\documentclass[%
 aip,
 jmp,%
 amsmath,amssymb,
 reprint,%
]{revtex4-2}

\usepackage{amsmath}
\usepackage{graphicx}
\usepackage{dcolumn}
\usepackage{bm}

\newcounter{inlineenum}
\renewcommand{\theinlineenum}{\roman{inlineenum}}
\newenvironment{inlineenum}
{\unskip\ignorespaces\setcounter{inlineenum}{0}%
	\renewcommand{\item}{\refstepcounter{inlineenum}{\textit{\theinlineenum})~}}}
{\ignorespacesafterend}

\begin{document}

\preprint{AIP/123-QED}

\title[]{Emergence of intelligent collective motion in a group of agents with memory}

\author{Danny Raj M}%
 \email{dannym@iisc.ac.in; dannyrajmasila@gmail.com}
\affiliation{Lab 10, Dept of Chemical Engineering, IISc Bangalore, Karnataka, India
}%

\author{Rupesh Mahore}%
\affiliation{Lab 10, Dept of Chemical Engineering, IISc Bangalore, Karnataka, India
}%


\date{\today}

\begin{abstract}
Intelligent agents collect and process information from their dynamically evolving neighbourhood to efficiently navigate through it.
However, agent-level intelligence does not guarantee that at the level of a collective; a common example is the jamming we observe in traffic flows.
%
In this study, we ask: how and when do the interactions between intelligent agents translate to desirable or intelligent collective outcomes? 
We explore this question in the context of a bidisperse crowd of agents with opposing desired directions of movement, like in a pedestrian crossing. 
We model a facet of intelligence, \textit{viz.} memory, where the agents remember how well they were able to travel in their desired directions and make up for their non-optimal past.
We find that memory has a non-monotonic effect on the dynamics of the collective.
When memory is short term, the local rearrangement of the agents lead to the formation of symmetrically jammed arrangements, which take longer to unjam.
However, when agents remember across longer time-scales, we find that the dynamics of an agent becomes sensitive to the relatively small differences in the history of the nearby agents. This gives rise to heterogeneity in the movement that causes agents to unjam more readily and form lanes that ease the movement. 
\end{abstract}

\keywords{Intelligent agents; memory; proportional integral controller; self organisation; Agent based models}
\maketitle


\section{Introduction}
A typical collective system comprises of agents that interact with each other through simple rules which lead to interesting, emergent group-level phenomena~\cite{Vicsek1995a, Reynolds1987, Couzin2002, Theraulaz2003}.
Generally, these phenomena can be classified as complex, since the features of the emergent behaviour are not explicitly encoded in the rules at the level of the agent and the observed behaviour is a result of the complex many-body interactions in the system~\cite{Sumpter2006}.
Examples include, the emergence of order from the inherent noise in collectives with small number of agents~\cite{Jhawar2020, Yates2009}; cohesion arising out of the choice of neighbours to interact with~\cite{Lei2020, Jadhav2022}; self assembly of pedestrian crowds into lanes following the steric interactions between agents in the system~\cite{Isobe2004, Kretz2006}; reaching a consensus via stigmergic interactions mediated through pheromones in an army of ants~\cite{Schmidt1992, Moussaid2009a}.
While the group-level behaviour is non-trivially connected to the agents and their interactions, it is also very interesting to note that these phenomena offer significant advantages to the agents when they are in a collective than when they are isolated. For instance, the coordinated motion in social animals helps the collective evade a predator better and also have an increased foraging efficiency~\cite{Romanczuk2009,Papadopoulou2022, Ioannou2012}; the formation of lanes in crowds eases the motion of pedestrians~\cite{Kretz2006, Feliciani2016}; ants optimise their travel times, and resource allocation via stigmergic interactions~\cite{Reid2015, Detrain2006}.
So it has always been of interest to understand how useful (intelligent) collective behaviours emerge from simple interactions at the level of the agents.

Agents that produce these intelligent group-level outcomes need not necessarily be intelligent, to begin with~\cite{Reid2016}. Even those agents with limited cognition, access to information and ability to process can collectively produce intelligent behaviour.
However, let us focus our attention on human collectives where the individual agents are intelligent and capable of sophisticated cognitive functions. Here, one could then ask: does the intelligence of human agents always guarantee intelligent collective outcomes? We know the answer to this question is a `no', just from our daily experience with vehicular traffic and pedestrian movement. Human collectives are capable of exhibiting both intelligent manoeuvres that increase throughput, and unintelligent collective behaviour like jamming~\cite{Tsuboi2020, Chauhan2021a}. The question then becomes: when do intelligent group-level behaviours emerge in a collective system? What aids or prevents this emergence?

To answer these questions, we explore the dynamics---more specifically the motion---of a collective composed of intelligent agents and study how their interactions lead to group level phenomena.
By investigating the phenomenology underlying this emergence, we identify when the group-level behaviour exhibited is intelligent and when it is not.
To model intelligence, one may have to simulate the cognitive process through which intelligent agents reach their decisions that affect the dynamics. Writing down a model for a brain is no simple task and will make interpreting the results extremely hard. Even an artificial neural network model would result in a large number of parameters that have to be tuned and explained.
Another way to the model would be, to use principles that agents broadly follow to make decisions, as a proxy for a model of the process of thinking. For example, one could use the principle of causal entropy or future state maximisation, where agents strive for states that offer better future possibilities~\cite{Hornischer2019,Charlesworth2019}.
However, current formulations of these models do not incorporate the `goal' of the agents and the competition between achieving the goal vs maximising the causal entropy~\cite{Wissner-Gross2013, Mann2015}.
Hence, we resort to the classic social force-based framework to model the interactions and movement of the agents. The agents have the desired goal defined by a desired speed and direction of motion. Agents always try to restore their state to the desired state.
Agent intelligence is modelled as a force affecting agent dynamics based on collected neighbourhood information.
Here, we take a minimalist approach in modelling a single aspect of intelligence, namely \textit{memory}. Agents in our study remember their past, they collect information on how well they were able to achieve their goals. The force arising due to the memory of the agent tries to restore the agent to its desired, making up for failed attempts in the past.
If the memory of the agents helps the collective achieve a state where the agents' individual goals are achieved swiftly, then we call the collective-state as `intelligent'. 

In this article, we first present the model for agents with memory. We demonstrate the dynamics using a bidisperse collective where there are two groups of agents with opposing desired directions of motion, meeting head-on. 
We explore the effect of memory on the dynamics of just a pair of agents and then proceed to investigate the effect on the collective.
In some regions of the parameter space, we observe the emergence of collective-level intelligence and in some cases where the intelligence of the agents has a negative effect on the collective dynamics. We probe into the phenomenology of agent dynamics to understand the origins of the observed collective phenomena. We then make connections to lane-less traffic flow conditions, which is often the case in densely populated countries.

\section{Model collective and the study}
In this study, we consider a bidisperse collective where there are two types of agents: one group desires to move in the positive x direction while the other in the negative x direction. 
We use a social force model, similar to refs~\cite{Helbing2000a, Reichhardt2018, Nabeel2022}, to simulate the motion of these agents on a two dimensional domain, with periodic boundary conditions in the x direction and bounded by walls in the y direction.
When the velocity is scaled with the desired speed of motion ($s_0$), time by the inertia ($\tau$), memory force by $s_0 \times \tau$, we can write down the equations governing the dynamics of the agents as shown in Eq~\ref{eqn:socialforce} and \ref{eqn:memoryforce}.

\begin{eqnarray}
    \label{eqn:socialforce}
	\frac{d\mathbf{v}_i}{dt} = (\mathbf{v}_{0, i} - \mathbf{v}_i) + \beta \mathbf{M}_i + \sum_{\forall j \neq i} \mathbf{F}_{ij} + \mathbf{F}_{i,w}\\
	\nonumber \\
	\label{eqn:memoryforce}
	\frac{d\mathbf{M}_i}{dt} = - \frac{\mathbf{M}_i}{\alpha} + (\mathbf{v}_{0, i} - \mathbf{v}_i)
\end{eqnarray}

The first two terms in the RHS of Eq \ref{eqn:socialforce} are the restitution and the memory forces, respectively. 
The restitution force drives an agent to its desired direction and speed at a rate proportional to the deviation of the agent's instantaneous velocity $\mathbf{v}_i$ from the desired $\mathbf{v}_{0,i}$. 
The memory force is an additional social force, coming from the agent's movement history, with its own time dynamics (see Eq~\ref{eqn:memoryforce}). The origins of this force is due to the cognitive faculties intrinsic to the `intelligent' agent.
The deviations from the desired velocity acts as the source for $\mathbf{M}_i$ while at a rate proportional to $\frac{\mathbf{M}_i}{\alpha}$, memory also decays. Here, $\alpha$ is the time scale corresponding to how far past the agent remembers.
The more the agent has slowed down or is pushed in a direction different from its desired, \textit{i.e.}, $\mathbf{v}_i \neq \mathbf{v}_0$, the more is the memory-force that makes the agent make up for its non-optimal history.
This feedback mechanism that the agent exhibits through memory is similar to how control architectures are implemented, where the control action depends on the deviation(s) from set point.
More information on the memory model and its connection with control theory can be found in an upcoming article as part of the Traffic and Granular flow 2022 conference proceedings.

Depending on whether an agent belongs to groups 1 or 2, it will have an appropriate $\mathbf{v}_0$, as given by Eq~\ref{eqn:Desdir}.
\begin{equation}
    \label{eqn:Desdir}
	\mathbf{v}_{0, i} =\left\{\begin{matrix}
		+ \mathbf{e}_x & i \in \text{Group 1}\\ 
		- \mathbf{e}_{x} & i \in \text{Group 2}
	\end{matrix}\right.
\end{equation}
Short-ranged repulsive forces (with cutoff distance of $l_{cr}$) are used to model both the inter-agent and the agent-wall interactions (see Eqs~\ref{eqn:interagforce}, \ref{eqn:wallforce}). 
\begin{eqnarray}
	\label{eqn:interagforce}
	\mathbf{F}_{ij} =\left\{ \begin{matrix} 
		-\gamma (d_{ij}-2)^{-3} \hat{\mathbf{d}}_{ij} & d_{ij}<l_{cr} \\
		0 & \text{\textit{otherwise}}
	\end{matrix} \right. \\
	\nonumber \\
	\label{eqn:wallforce}
	\mathbf{F}_{iw} =\left\{ \begin{matrix} 
		-\gamma (\|\text{wall} - y_i\|-1)^{-3} \hat{\mathbf{d}}_{ij} & d_{ij}<l_{cr} \\
		0 & \text{\textit{otherwise}}
	\end{matrix} \right.
\end{eqnarray}
These make sure that the agents compete for space as they move past each other. The dynamics observed does not sensitively dependent on the choice of the interaction model employed.

\paragraph*{The study:}
The agents from the two groups are placed on two sides of the domain and are allowed to meet head-on (as shown in the schematics of figures~\ref{fig:twoagentdynamics} A and \ref{fig:crowddynheadon} A). These agents, as they are driven in their desired directions, will collide with agents from the other group and slow down. This deviation from the desired velocity will cause an increase in the memory force which will help the agent force its way through, even more. The agents will form a temporarily jammed state which will eventually unjam to give rise to a laned configuration that allows for free motion of the agents.

The objective here is to understand how memory---specifically the strength of the social force $\beta$ and the time scale $\alpha$---affects the crowding dynamics. When $\beta = 0$, memory has no effect on the motion of the agents. When $\alpha$ is low, the agent dynamics is affected only by the information from its recent past and similarly when $\alpha$ is high the agent dynamics is affected by a longer history.

First, we explore the dynamics of two agents meeting head-on, as a function of $\beta$ and $\alpha$. Then, we study the dynamics of the collective (40 agents; 20 in each group) as a function of the memory parameters.

\section{Results and Discussion}
\subsection{Effect of memory on the head-on collision of two agents.}

\begin{figure*}
    \centering
    \includegraphics[width=\linewidth]{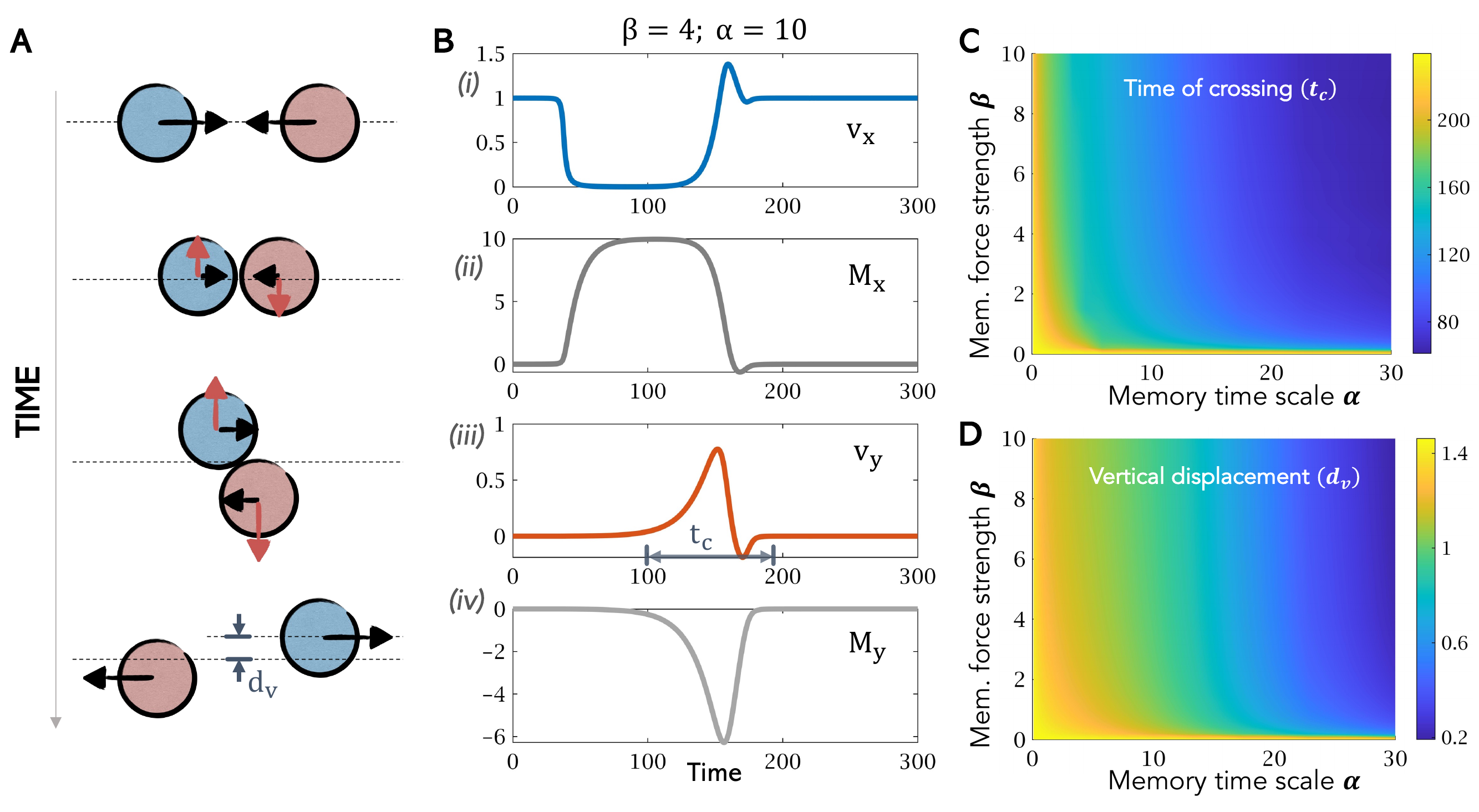}
    \caption{\textbf{Dynamics of two agents meeting head-on}. \textbf{A} - Schematic of the two agents meeting head-on with arrows marking their velocities in both x and y directions separately. Agents slide past each other to a final configuration where agents are displaced vertically $d_v$ from their original $y$-positions. \textbf{B (i - iv)} - Time evolution of velocity and memory in both x and y directions for a single agent. The time of crossing ($t_c$) is marked in (iii), since it is based on the condition: $v_y>10^{-3}$. \textbf{C} - Time of crossing ($t_c$) heat map as a function of the memory parameters $\alpha$ and $\beta$. \textbf{D} - Vertical displacement ($d_v$) heat map as a function of $\alpha$ and $\beta$.}
    \label{fig:twoagentdynamics}
\end{figure*}

To understand how memory affects the crowding dynamics we begin our analysis by looking at the dynamics of two agents: one from each group, meeting head-on (schematic in 
figure~\ref{fig:twoagentdynamics} A).
The agents are far away from the walls and the only interaction they experience is the inter-agent force. This is directed along the line joining the centres of the agents which slows the agents down when they approach each other. Since, agents are not present along the same horizontal axis (perturbed in the vertical direction), they experience a small y-component force that pushes them in the vertical direction. The agents eventually slide past each other occupying separate lanes.
In the presence of memory, this slowing down results in the increase of the memory force which in turn affects the motion of the agents and speeds up the dynamics (see figure~\ref{fig:twoagentdynamics} B (i, ii) and (iii, iv)).

The memory force plays two roles. Firstly, as $M_x$ increases, it quickens the time taken for the agent to begin sliding over the other agent. Now, as the agent slides over, it moves in the y direction which is not desired (as far as the goal of the agent is considered). Hence, $M_y$ starts to increase in a negative direction. However, since the inter-agent force dominates the sliding process, $M_y$ does not have a role in the motion of the agent during that time. But once, the sliding is over, the inter-agent force begins to reduce in magnitude and now the memory force in y, $M_y$, gives rise to a negative $v_y$ which attempts to make up for the vertical displacement experienced by the agent during the sliding process.

Hence, to characterise the dynamics we compute two parameters: 
\begin{enumerate}
    \item \textit{Time of crossing} $t_c$, which is the time taken by the agents to complete the sliding process. We define this based on the total time $v_y$ was greater than $10^{-3}$ (marked in figure~\ref{fig:twoagentdynamics}B (iii)) and,
    \item \textit{Vertical displacement} $d_v$ which is the displacement of the agent in the y direction from its initial y position (marked in figure~\ref{fig:twoagentdynamics}A).
\end{enumerate}
Figures~\ref{fig:twoagentdynamics} C and D show the heat maps of $t_c$ and $d_v$ for the memory parameters $\alpha$ and $\beta$. We observe that, in this simple two-agent case, memory has a monotonic effect on the dynamics. The higher the effect of memory, either through a longer time scale $\alpha$ or by a larger force strength $\beta$, the quicker the sliding and lower the displacement. In fact, when $\alpha$ is very high $(>20)$, we see that the agent positions are almost fully restored in the y-direction.

\subsection{Effect of memory on the dynamics of the collective.}
\begin{figure*}
    \centering
    \includegraphics[width=\linewidth]{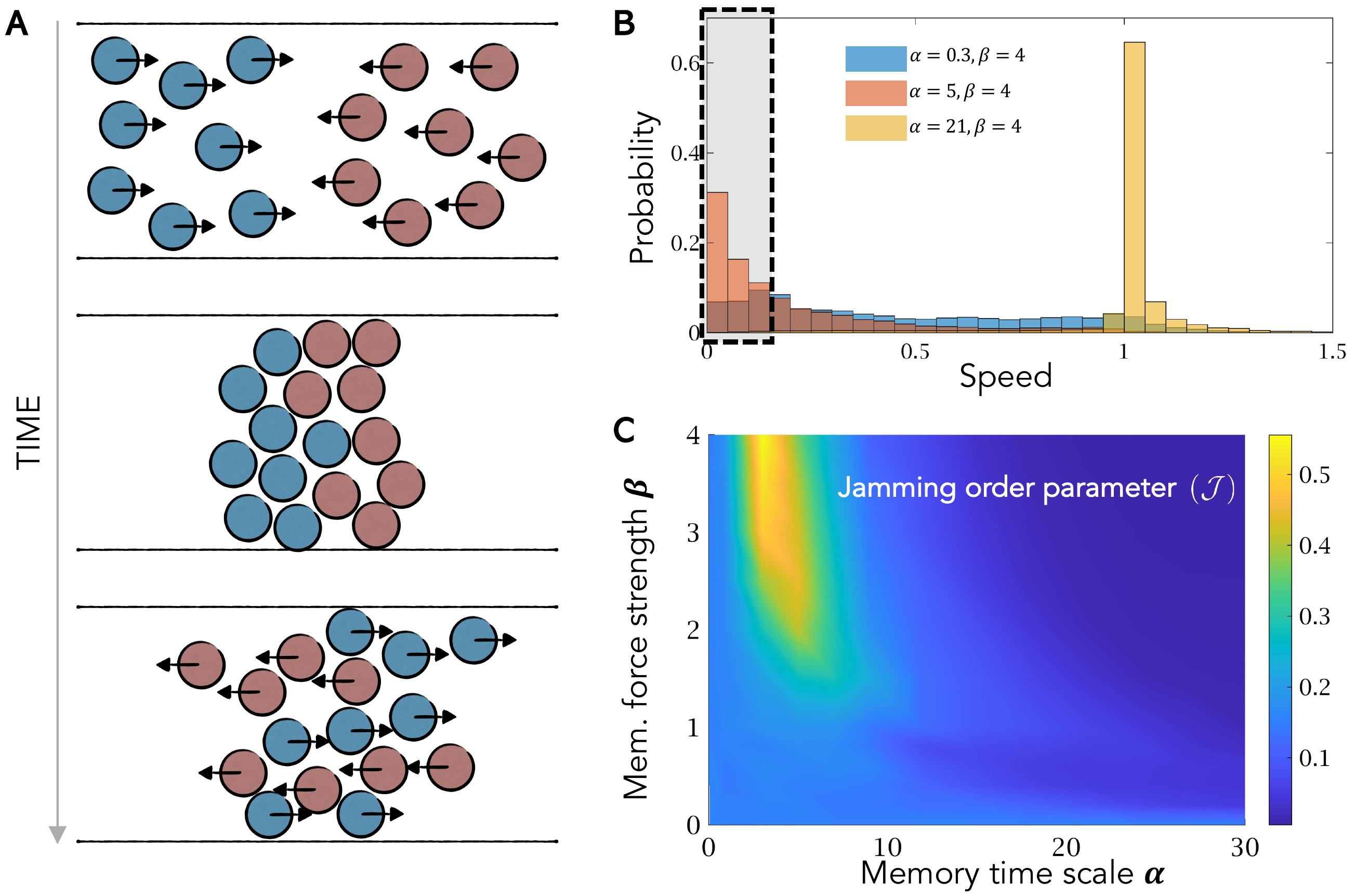}
    \caption{\textbf{Dynamics of a bidisperse collective meeting head-on}. \textbf{A} - Schematic of the dynamics of the agents in a 1D periodic domain in x and bounded by walls in y. \textbf{B} - Probability of finding an agent of a given speed, plotted for three cases: $\alpha = 0.3, \beta = 4$, $\alpha = 5, \beta = 4$ and $\alpha = 21, \beta = 4$. \textbf{C} - The order parameter $\mathcal{J}$ heat map as a function of the memory parameters $\alpha$ and $\beta$.}
    \label{fig:crowddynheadon}
\end{figure*}

Next, we simulate a collective of 40 agents, with 20 in each group. Agents of group 1 are placed on the left-hand side of the domain and group 2 on the right (as shown in schematic figure~\ref{fig:crowddynheadon} A). As the system evolves in time, agents meet head-on near the centre of the domain where they form a compact arrangement,a jammed configuration.
A variety of different scenarios are possible here, from a few agents in a jammed configuration to the entire collective at a standstill.
Following this, the agents unjam dynamically leading to the formation of lanes that ease the movement of the agents thereafter. 

To understand how memory impacts the movement of agents in the crowd, we use a jamming order parameter $\mathcal{J}$ which is defined as the probability of finding a slowly moving agent. If many agents move slow most of the time, $\mathcal{J}$ will take a large value and it would mean that collisions and local jamming are slowing down these agents considerably. At the same time, if $\mathcal{J}$ is small, it would mean that the agents were able to escape getting jammed and were able to move as desired.
Here in our study, we call an agent slow if it moves with a speed less than $15\%$ of the desired speed $\|\mathbf{v}_0\|$. Figure~\ref{fig:crowddynheadon} B shows the histograms of the speeds for a few realisations for different values of memory. The boxed region identifies the probability of finding an agent moving with a speed $<0.15$.

We compute $\mathcal{J}$ for different values of the memory time scale $\alpha$ and strength of force $\beta$ (see figure~\ref{fig:crowddynheadon} C). 
Interestingly, we find that memory has a non-monotonic effect on crowd dynamics. When the agent remembers only its recent past, \textit{i.e.} the memory is short-term $0< \alpha < 8$ (and with sufficiently large $\beta$), we find that agents have a large propensity to form jammed configurations that take much longer to unjam giving rise to large values for the order parameter $\mathcal{J}$.
At the same time, we also find that for large enough $\alpha$, the movement dynamics become very efficient taking very low values of $\mathcal{J}$.

\begin{figure}
    \centering
    \includegraphics[width=1\linewidth]{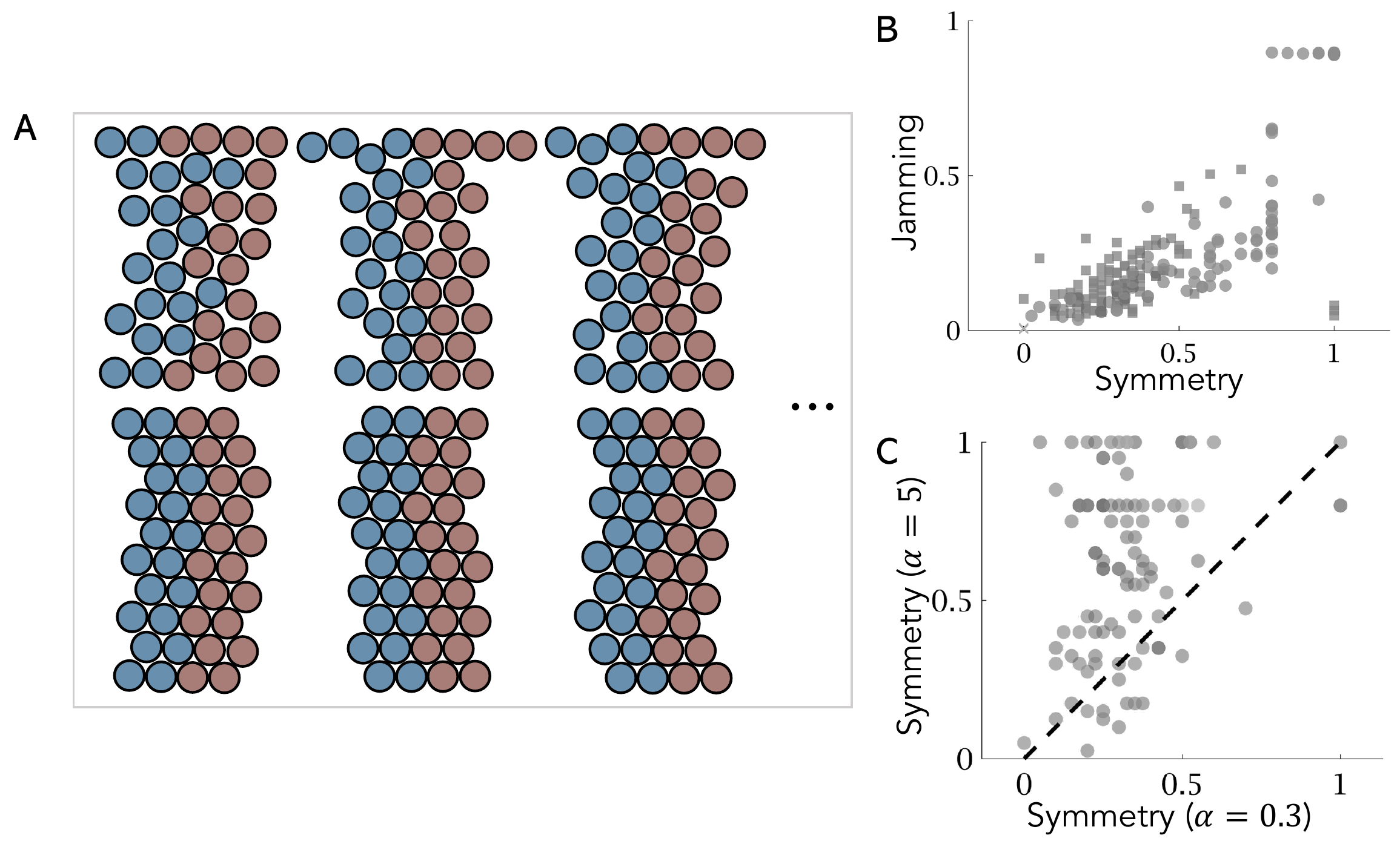}
    \caption{\textbf{A} - Jammed configurations formed for the same initial conditions for when memory is very small ($\alpha = 0.3$ and $\beta = 4$), and when memory is short term ($\alpha = 5$ and $\beta = 4$). \textbf{B} - Jamming order parameter $\mathcal{J}$, plotted as a function of the symmetry order parameter, which is the fraction of layers that were symmetric. $(\Box)$: corresponds to $\alpha=0.3$, $(\bigcirc)$: $\alpha=5$ and $(\times)$: $\alpha=21$ (only found at the origin). \textbf{C} - Symmetry order parameter plotted for both $\alpha = 0.3$ and $\alpha = 5$. The dotted line is the $45^{\circ}$ line. Every point in the plots shown in B and C corresponds to a single realisation of the agent dynamics, which corresponds to one set of initial positions and velocities of agents.}
    \label{fig:configs_shorttermmemory}
\end{figure}

\subsubsection{Short-term memory causes the self-assembly of symmetric interlocks.}
To understand why short-term memory gives rise to a high $\mathcal{J}$, we look into the structure of the jammed configurations formed when agents meet head-on, for different values of the memory parameters.
We find that short-term memory gives rise to jammed configurations that are symmetric,\textit{i.e.}, there are an equal number of agents from either group in a single lane (as shown in figure~\ref{fig:configs_shorttermmemory} A).
Since the forces are more or less balanced for such assemblies, it takes much longer to unjam giving rise to a high value for $\mathcal{J}$. Figure~\ref{fig:configs_shorttermmemory} B shows that $\mathcal{J}$ increases with the symmetry of the jammed configuration. Here, symmetry is simply a measure of the fraction of layers in the jammed configuration that have the same number of agents from groups 1 and 2.

Interestingly, we find that even when the initial conditions (x and y positions at $t=0$ for all agents) are the same, short-term memory aids in the formation of symmetric assemblies when those formed with very low levels of memory are asymmetric. Figure~\ref{fig:configs_shorttermmemory} C shows the scatter plot of the symmetry measure for the $\alpha=5$ (short-term) and $\alpha=0.3$ cases. Most points lie above the diagonal line, which clearly indicates that the rearrangements facilitated by the memory force term gave rise to more symmetric arrangements.

When agents meet head-on, they have to slide past each other. Now, when the effect of memory is low, the time taken to cross each other is high. We know this from the $t_c$ computed from our two-agent simulations (figure~\ref{fig:twoagentdynamics} C). As the agents in the collective begin to crowd near the centre of the domain
to form a compact arrangement, very soon the space needed for crossing is no longer available. Hence, significant rearrangements seldom happen in the absence of memory. However, as $\alpha$ increases, the time for crossing $t_c$ reduces. This results in possible rearrangements as the agents form the jammed configuration, to yield symmetric interlocks.

\begin{figure}
    \centering
    \includegraphics[width = 1 \linewidth]{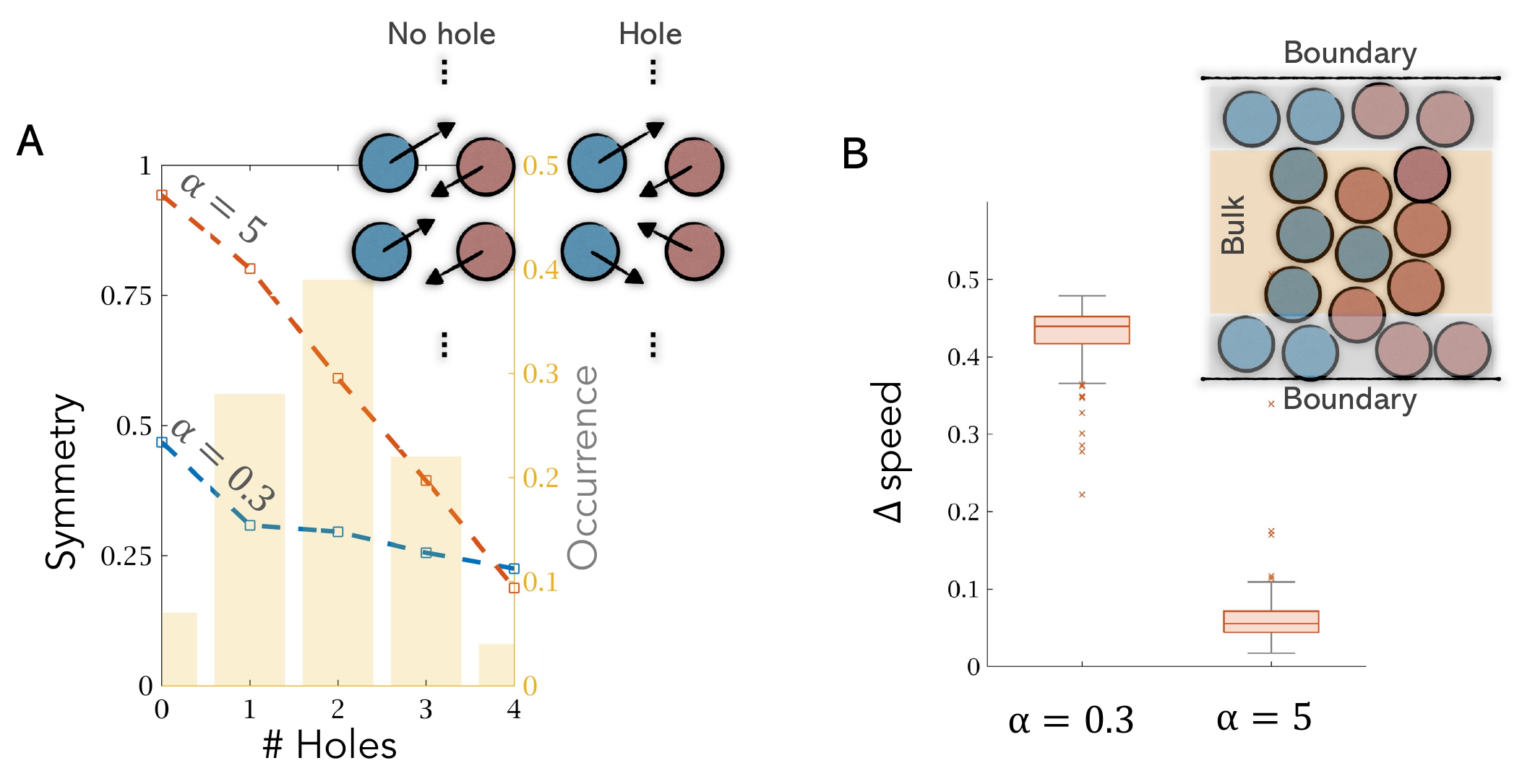}
    \caption{\textbf{A} - The mean layering order parameter that quantifies the symmetry is plotted against the number of holes formed in the front of the assembly. The bar plots in the background (with the axis in the right) represent the probability of occurrence of these holes the different realisations of the agent dynamics. INSET: schematic of assemblies with and without a hole.  \textbf{B} - The difference in the average speeds between the bulk of the assembly and the agents near the boundary is plotted for $\alpha = {0.3, 5}$. INSET: schematic showing the bulk and the boundary.}
    \label{fig:originofasymmetry}
\end{figure}

\paragraph{Origins of symmetry through the formation of holes.}
To understand the phenomenology behind the assembly of compact jammed structures, we investigate the dynamics near the interacting fronts of the two groups.
As the fronts come into contact, agents of one group have to slide past those in the other in front of it, in order to move in their desired directions.
This gives rise to interesting features in the self-assembly of agents in that region which we believe dictates the dynamics and eventually the final state of the compact jammed arrangement.

When agents approach each other, their relative positions would result in the pair sliding either in a clockwise or an anticlockwise fashion. If two adjacent pairs slide in the same manner, both of them clockwise or anticlockwise, the agents rearrange in a rather symmetric fashion. The agents behind these fronts also follow this reorganisation giving rise to symmetric jammed configurations, esp. when memory is short-term.
At the same time, when adjacent pairs of agents slide in opposite directions, one anticlockwise and the other clockwise, the agents open up as they move away from each other creating a \textit{hole}. Now the nearby agents compete for the space created. When memory is short-term, all agents in the interacting front reorganise in a similar fashion which essentially amounts to crowding. This leads to the emergence of local asymmetry in the assembly of agents.

The inset in figure~\ref{fig:originofasymmetry} A shows the schematic of the local movement of agents that may or may not give rise to a hole. 
As expected, we find the symmetry of the jammed configuration to decrease as the number of holes in the assembly increases (as shown in figure~\ref{fig:originofasymmetry} A). Here, the formation of holes can be predicted from the y positions of the agents at $t=0$.
The number of holes formed in any given realisation depends on the different ways in which agents at the interacting front move. Only when all the interacting pairs move in a similar fashion we get 0 number of holes. However, even if a single pair moves in another direction, in the bulk, it would result in 2 holes and if that is in the boundary, 1 hole and so on. Hence, the probability of the occurrence of a hole depends on the number of ways in which one can achieve them. The bar plot in the background of figure~\ref{fig:originofasymmetry} A shows the propensity to get a certain number of holes in the assembly.

\paragraph{Origins of asymmetry through boundary effects.}
In addition to the dynamics dictated by the formation of holes, we also observe the boundary playing a role in the origin of asymmetry in the jammed configuration. Agents in the boundary have to slide past those on the one side while on the other, they are constrained by the immovable boundary.
When the effect of memory is very low, we find that agents near the boundary are unable to participate in the dynamics which results in the reorganisation in the bulk. They begin to participate later and when they do so, they result in asymmetric arrangements. The inset of figure~\ref{fig:originofasymmetry} B shows schematic of the agents in the boundary and the bulk. We compute the difference in the speeds of agents in these two regions (during the initial transients) and we find that $\alpha = 0.3$ has a higher difference than when the memory is short-term, $\alpha = 5$.

\begin{figure}
    \centering
    \includegraphics[width = 0.7 \linewidth]{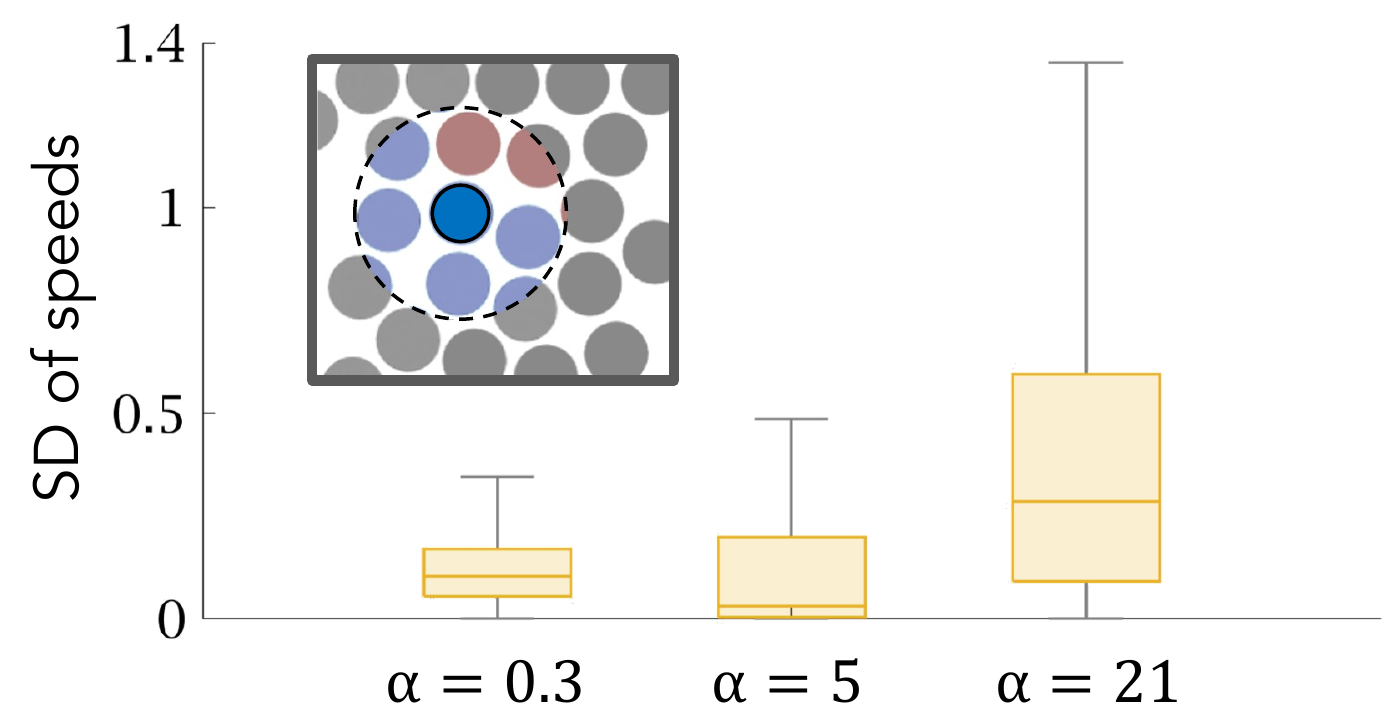}
    \caption{Box plots of the standard deviation of speeds evaluated around the neighbourhood of every agent for three cases: $\alpha = 0.3$, $\alpha = 5$ and $\alpha = 21$ and $\beta = 4$. INSET: shows how the neighbours are computed around every agent.}
    \label{fig:highalpha_heterogeneity}
\end{figure}

\subsubsection{Long-term memory leads to the heterogeneity in movement that aids in unjamming.}
We observe that when the memory is long-term, \textit{i.e.} $\alpha$ is large, agents form lanes very effectively and rapidly giving rise to small values of $\mathcal{J}$.
To form lanes, the agents have to slide past each other quickly before the other agents crowd the area. We know from our two-agent study that $t_c$ is reduced considerably when $\alpha$ is high. In addition, we also know that agents restore the vertical displacement $d_v$, which results in the effective utilisation of space as agents form lanes.
When holes are present, we find that it facilitates the assembly of multiple layers of agents from the same group. 

To understand how agents efficiently layer at the level of the collective, we look at the distribution of agent speeds in the neighbourhood of every agent. We compute the standard deviation of the speeds of those neighbouring agents that are within 3 times the radius of the focal agent (as shown in the inset of figure~\ref{fig:highalpha_heterogeneity}).
We find that there is a large variation in the speeds of the agents when $\alpha$ is high.
Since, the memory parameters are the same for all the agents, the general expectation is that they will experience similar amounts of the memory force as they crowd and slow down. However, when $\alpha$ is large, we find that memory of an agent is sensitive to the velocity changes it experiences. Hence, small differences in the history of the agent movements give rise to large differences in the memory forces giving rise to heterogeneous movement locally and a corresponding large variation in the speeds of the agents. 
Therefore, when a couple of agents slide past each other, the neighbouring agents do not make appreciable movement allowing them to layer efficiently.

\section{Conclusion}
\paragraph*{Summary:}
In this article, we explore how and when agent-level intelligence translates to intelligent group-level behaviour in collective systems. Using a bidisperse group of agents (with two opposing desired directions) that possess some facet of intelligence, \textit{viz.} memory, we study the emergence of intelligence at the level of the collective. 
Although memory guarantees efficient movement of the agents when there are only two of them, we find that it can give rise to both efficient movement and in some cases sub-optimal jamming behaviour, when there are enough agents to span the width of the domain.
We show that the cause for the diverse behaviours is due to:
\begin{inlineenum}
    \item the time scales involved in crowding as agents approach each other to form compact arrangements, 
    \item the time-scales associated with the rearrangement processes that lead to agents of one group sliding past another to escape a local jamming event and
    \item the presence of the boundary.
\end{inlineenum}
In the absence of memory, agents do not have the time required to move past other agents, esp. near the boundary, and they end up in jammed configurations. However, these configurations are highly asymmetric and eventually unjam. However, if the agents possess some memory, \textit{viz.} short-term, they have a large propensity to form symmetric interlocks. The memory facilitates the rearrangement of all the agents in a similar fashion, which result in crowding and assembly into symmetric arrangements.
Now, when the effect of memory is large, \textit{i.e.}, agents remember over longer time scales, we find that small differences in the movement between agents result in large differences in the memory force that gives rise to locally heterogeneous motion of the agents which facilitates the formation of lanes.

\paragraph*{Connections to dense, lane-less traffic:}
These results are particularly interesting from the context of dense, lane-less traffic flows like that in countries like India, where we find intelligent agents as the cause for both unintelligent interlocks that reduce the efficiency of the collective and complex manoeuvres that increase the throughput of traffic in dense flow conditions.
Each vehicle on the road can be thought of as a selfish agent that is optimising or adjusting its motion to achieve its objective better. 
The agents used in this study too behave as controllers, specifically of the Proportional-Integral type, which uses information pertaining to its own past movement (memory) along with its current state (restitution), to achieve its set-point (\textit{i.e.} its desired velocity) effectively.
This similarity allows us to derive some general insights into the traffic flow problem from our study.
We show that when all the agents selfishly adjust to achieve their own objective, they can form symmetric interlocks that bring the movement to a standstill. Here, symmetry is crucial as agents are of the same type and size and they exhibit an inter-agent social force. However, in real traffic the interlock need not be symmetric; as long as the agents form arrangements where they oppose the motion of the other and prevent sliding past one another, jammed configurations will be formed.
This phenomenon is also similar to the faster-is-slower effect observed in pedestrian systems where throughput decreases when all agents trying to exit a room do so aggressively~\cite{Helbing2000,Garcimartin2014}.
At the same time, when conditions favour heterogeneous movement of agents, \textit{i.e.}, some agents move faster in comparison to others, we show that it is possible for agents to achieve effective movement.
This translates to an efficient use of the available space in a limited time, which is often the mechanism behind efficient manoeuvring seen in real traffic, where some agents might move faster than others resulting in a rather asynchronous use of space that might bypass a jamming event. 
We show that both these types of movements emerge in a collective of agents with intelligence in the form of memory. 

\begin{acknowledgments}
DRM thanks Prof Kumaran, IISc for the discussions during the initial stages of the project. The authors thank Arshed Nabeel for the careful reading of the manuscript. The authors thank DST INSPIRE faculty award (grant number: DST/INSPIRE/04/ 2017/002985) for the funding.
\end{acknowledgments}



\section*{References}
\bibliography{references.bib}

\end{document}